\documentclass[11pt]{article}
\usepackage{graphicx,verbatim,array,multicol,palatino,amssymb,amsfonts, amsmath, subfigure}
\usepackage{color}
\usepackage{psfrag}
\usepackage{RM}
\def\bx{{\bf x}}

\def\lboxit#1{\vbox{\hrule\hbox{\vrule\kern6pt
      \vbox{\kern6pt#1\kern6pt}\kern6pt\vrule}\hrule}}

\def\thick#1{\hbox{\rlap{$#1$}\kern0.25pt\rlap{$#1$}\kern0.25pt$#1$}}

\def\bbeta{{\thick\beta}}

\def\by{{\bf y}}
\def\bu{{\bf u}}

\def\bX{{\bf X}}
\def\bZ{{\bf Z}}

\def\bI{{\bf I}}

\def\bSig{{\bf \Sigma}}
\def\blam{\underline{\lambda}}
\def\bx{{\bf x}}
\def\bX{{\bf X}}

\def\bA{{\bf A}}
\def\bI{{\bf I}}

\def\jump{\vskip3mm\noindent}

\newtheorem{prop}{Proposition}

\begin{document}
\null\vskip2cm
\begin{center}

{\LARGE
\bf Adaptive Optimal Scaling of Metropolis-Hastings Algorithms Using the Robbins-Monro Process}
\vskip5mm
\begin{large}
{\sc P. H. Garthwaite} \\
{\it Department of Mathematics and Statistics, Open University,\\
Milton Keynes, MK7 6AA, U.K.}
\vskip3mm

{\sc Y. Fan}\\
{\it School of Mathematics and Statistics,\\
University of New South Wales, Sydney, 2052, AUSTRALIA}

\vskip3mm
{\sc S. A. Sisson }\\
{\it School of Mathematics and  Statistics,\\
University of New South Wales, Sydney, 2052, AUSTRALIA}

\end{large}
\vskip5mm

{\sc Abstract}\vskip2mm
\end{center}
We present an adaptive method for the automatic scaling of Random-Walk Metropolis-Hastings
algorithms, which
quickly and  robustly identifies the scaling factor that yields a specified overall sampler acceptance probability.
Our method relies on the use of the Robbins-Monro search process, whose performance is determined by an unknown steplength constant. We give a very simple estimator of this constant for proposal distributions that are univariate or multivariate normal, together with a sampling algorithm for automating the method. The effectiveness
of the algorithm is demonstrated with both simulated and real data examples.
This approach could be implemented as a useful component in more complex adaptive Markov chain Monte Carlo algorithms, or as part of automated software packages.

\jump
\noindent
{\em Some keywords}: Adaptive Markov chain Monte Carlo; Robbins-Monro; optimal scaling; random-walk Metropolis-Hastings.

\vskip4mm

\section{Introduction}

Markov chain Monte Carlo (MCMC) algorithms are now routinely used in Bayesian statistical inference. In
particular, the Metropolis-Hastings algorithm \shortcite{metropolis+rrtt53,hastings70}
is highly
popular due to its simplicity and general applicability. The most commonly implemented variant is the Random-Walk Metropolis-Hastings sampler (RWMH). The RWMH sampler uses a proposal distribution (most commonly,
the Gaussian distribution) centered on the current value of the Markov chain,
with some specified scale parameter $\sigma^2>0$. The success and efficiency of RWMH depends
heavily on the value of the scale parameter, which typically produces a smaller acceptance probability for a proposed move  when it is large, and a larger acceptance probability when it is small.

For target distributions of a certain form,
\shortciteN{roberts+gg97} proved
that an optimal acceptance rate should be 0.234 based on a multivariate proposal distribution. \citeN{roberts+r01}
suggest a value of 0.44 for a univariate proposal distribution. Though these theoretical results were developed under fairly restrictive assumptions, they have proven to
work well for more general problems. Thus a common practice in MCMC is to manually tune acceptance probabilities by varying the scale parameter to achieve the above rates.
In practice, the manual tuning of scale parameters becomes difficult when the number of proposal distributions in the MCMC sampler is large, and can be further complicated by correlation between parameters.
One should also avoid asking users to tune algorithm parameters,
when developing software
for general (non-expert) use.

In this article
we consider the use of a stochastic search algorithm -- the Robbins-Monro process -- to
automatically tune the scale parameters. In essence, the
resulting adaptive sampler will
increase $\sigma$ if the previous MCMC proposed move was accepted and decrease $\sigma$ if the proposal was
rejected. The amount by which $\sigma$  is changed (the step size) decreases linearly with the number of iterations in the Markov chain. Thus this adaptation procedure satisfies the {\it diminishing adaptation
criterion} (\citeNP{roberts+r09,rosenthal10}). In addition to the diminishing
adaptation condition, any adaptive scheme also has to satisfy the {\it containment} (or bounded convergence) condition \shortcite{bai08} in order to preserve the ergodicity of the Markov chain.  This technical condition is satisfied for virtually all reasonable adaptive schemes \cite{rosenthal10}, and is satisfied in our context since the step size will approach zero. 
The rate of convergence of the Robbins-Monro process depends on a steplength constant that controls the magnitude of the step size. Estimation of this constant is the focus of this paper.

\citeN{andrieu+t08} review a variety of adaptive MCMC methods, including several that use variants of the Robbins-Monro process. None of the algorithms they describe attempt to estimate the optimal value of the steplength constant. One co-author of the review had proposed a method of estimating this constant \cite{andrieu+r01}, but a quantity that is critical to the method could only be estimated by combining information from three separate Markov chains, making the accuracy of the estimate questionable. The method is not mentioned in \citeN{andrieu+t08}.
See e.g. \citeN{roberts+r09}, \shortciteN{haario+st05},
\shortciteN{craiu+ry09}, for other recent work on adaptive MCMC methods.

In Section \ref{sec:RM}, we introduce the Robbins-Monro process, and its use within the context of the MCMC framework. Section \ref{sec:est} details results on estimation
of the optimal steplength constant for the Robbins-Monro process, while Section \ref{sec:example} provides the recommended adaptive optimal scaling algorithm. Simulated and real data analyses are performed in Section \ref{sec:examples}. We conclude with some discussion in Section \ref{sec:disc}.

\section{The Robbins-Monro process}

\label{sec:RM}
The standard situation for which the Robbins-Monro process was devised is the following. A binary response has probability of success $p(\sigma)$, where $\sigma$ is a parameter that can be controlled. It is assumed that $p(\sigma)$ is a monotonic function of $\sigma$ and here it is appropriate to suppose that the function is monotonically {\em decreasing}. This assumption usually holds in Random-Walk Metropolis-Hastings (and we make this assumption here), as a smaller scale parameter $\sigma$ generally corresponds to a larger acceptance rate $p(\sigma)$, and vice versa.  The aim is to find the value of $\sigma$ that gives a specified probability of success, say $p^*$. Let $\sigma ^*$ denote this value, so that $p(\sigma ^* ) = p^*$. The Robbins-Monro process conducts a stochastic search in which a sequence of trials is implemented. At each trial  $\sigma$ is set equal to the current estimate of $\sigma ^*$.  If the result of the trial is a success then the estimate of $\sigma^*$ is increased and, if it is a failure, the estimate is reduced. Let $\sigma _i$ denote the estimate of $\sigma^*$ at the $i$-th trial, $i=1,2,\ldots$.
The Robbins-Monro process (\shortciteNP{robbinsm54}) steps from $\sigma_i$ to  $\sigma _{i+1}$ according to the rule
\[
\sigma _{i+1} = \left\{ \begin{array}{ll}
\sigma _i + c(1-p^*)/i & \mbox{ if the $i^{th}$ trial is a success}  \\
\sigma _i - cp^*/i & \mbox{ if the $i^{th}$ trial is a failure,}
\end{array} \right.
\]
where $c>0$ is a chosen constant.
Each value of $\sigma_i$ is expected to be closer to $\sigma^*$ than the preceding one
(\shortciteNP{garthwaite+b92}). The size of steps is controlled by $c$, commonly referred to as the {\em steplength} constant.

The optimum choice of the steplength constant is $ c^* = -1/[dp(\sigma)/d\sigma ]_{\sigma =\sigma^*}$, where the derivative is evaluated at the target value $\sigma =\sigma^*$ (\shortciteNP{hodgesl55}).
The method has good asymptotic properties (\shortciteNP{hodgesl55}; \shortciteNP{wchwabew96}; \shortciteNP{wetherill63}). In particular, as $i \rightarrow \infty$, $\sigma_i - \sigma ^*$ is asymptotically normally distributed with a mean of zero and a variance of $p^* (1-p^*)c^2 c^*/i(2c-c^*)$, provided that $c>c^*/2$. If $c=c^*$ is set equal to its optimal value, then the asymptotic variance of $\sigma_i$ equals the Cramer-Rao lower bound to the variance of any non-parametric unbiased estimator of $\sigma ^*$ \shortcite{wetherill75}. Moreover, as noted by \citeN{wetherill63}, the asymptotic variance is relatively insensitive to the precise value chosen for $c$, especially if $c$ overestimates $c^*$ so that steps are larger than their optimal size: the variance is one-third greater than its lower bound when $c = 2c^*$ or $c= 2c^*/3$.
In general, the optimal value $c^*$ is not known and must be estimated.

In the context of the Metropolis-Hastings algorithm, suppose
the posterior target distribution is $f(\bx) \propto f^{\#}(\bx)$, where $f^{\#}(\, .\,)$ is known. Let $g(\cdot \, |\, \bx,\sigma)$ be the proposal distribution when currently at $\bx$, where $\sigma$ is a scale parameter to be identified using the Robbins-Monro process.
Define $p(\bx, \sigma)$ to be the probability of accepting a proposed move from $\bx$, generated from $g(\cdot \, |\, \bx,\sigma)$.  We assume that for any $\bx$, $p(\bx, \sigma)$ is a monotonic decreasing function of $\sigma $. That is, we assume that the acceptance probability does not increase as the variance of the proposal distribution increases. Under the above scenario, a trial consists of generating a value from the proposal distribution. Accepting the proposed value equates to a success; not accepting it to a failure. When at $\bx$, the probability of success in the Robbins-Monro process, $p(\sigma)$,  is given by  $p(\bx, \sigma)$, the acceptance probability in the Metropolis-Hastings algorithm.
However, as $\bx$ varies over the target distribution, and because $p(\bx, \sigma)$ varies with $\bx$ as well as $\sigma$, we define $\rho(\sigma)$ as the overall acceptance probability (OAP) of the sampler, $p(\sigma)=\int p(\bx, \sigma)\, f(\bx)\, d\bx $. The aim is then to find a value $\sigma=\sigma^*$,  for which the OAP has some specified value, $\rho(\sigma)=p^*$.

Specifically, we have that
\begin{equation*}
p(\bx,\sigma) = \int \mbox{min}\left\{ \frac{f^{\#}(\by)\, g(\bx \, |\, \by,\sigma)}{f^{\#}(\bx) \, g(\by \, |\, \bx,\sigma)},\,1 \right\} g(\by \, |\, \bx,\sigma)\, d\by.
\end{equation*}
In the case of a symmetric proposal, $g(\by \, |\, \bx,\sigma)= g(\bx \, |\, \by,\sigma)$ for all $\bx$, $\by$, $\sigma$, and
\begin{equation}\label{eqn:p}
p(\sigma) = \int\int \mbox{min}\left\{ \frac{f^{\#}(\by)}{f^{\#}(\bx) },\,1 \right\} \, g(\by \, |\, \bx,\sigma)\, f(\bx) \, d\by \, d\bx.
\end{equation}
Hence, under standard regularity conditions
\begin{equation}\label{dpds}
\frac{dp(\sigma)}{d\sigma} = \int \int \mbox{min}\left\{ \frac{f^{\#}(\by)}{f^{\#}(\bx) },\,1 \right\}  \left( \frac{dg(\by \, |\, \bx,\sigma)}{d\sigma} \right) \, f(\bx) \, d\by \, d\bx.
\end{equation}
In this article, we restrict attention to the case where $g(\by \, | \, \bx, \sigma)$ is a univariate or multivariate normal distribution with mean $\bx $ and variance $\sigma ^2 \bA $, and so $g(\by \, |\, \bx,\sigma)= g(\bx \, |\, \by,\sigma)$.
The quantity $dp(\sigma)/d\sigma$, evaluated at $\sigma^*$, determines the optimal value of the steplength constant.
For the moment we assume that the covariance matrix $\bA$ is known. However, we relax this in Section \ref{sec:searchfull}, where we simultaneously estimate $\bA$ along with $\sigma^*$ within a standard adaptive random walk Metropolis algorithm (\shortciteNP{haario+st01}).

However, even in the usual Robbins-Monro context
 it is difficult to estimate the steplength constant from variation in $p(\sigma)$ (\shortciteNP{wetherill63}, \shortciteNP{ruppert91}). In the present context, attempting to estimate $c^*$ from variation in $p(\bx,\sigma)$
 is orders of magnitude harder, as $p(\bx,\sigma)$ is as sensitive to change in $\bx$ as to change in $\sigma$. In the following section we develop a procedure that avoids this problem.

\section{Estimation of the steplength constant}
\label{sec:est}

\citeN{garthwaite+b92} and \citeN{garthwaite96} provide an algorithm for finding confidence limits in Monte Carlo tests, in which the
steplength constant is not estimated through variation in $p(\sigma)$. Instead, the estimate of $c^*$ is based on the distance between the current estimate of one endpoint of the confidence interval and the point estimate of the quantity for which the interval is required. The ratio of this distance to the optimal steplength constant, $c^*$, is reasonably similar across a broad range of distributions -- sufficiently similar to provide an adequate estimate of the steplength constant. For the Metropolis-Hastings algorithm, results in this section suggest that, in a similar fashion, the relationship between $c^*/\sigma^*$  and $p^*$ may be sufficiently similar across distributions for $c^*$ to be adequately estimated from $\sigma^*$ and $p^*$. Propositions 1--3 below motivate the estimator of $c^*$ that we propose for univariate target distributions. Propositions 2 and 3 are proved in an appendix.

\begin{prop}
\label{PropBound} Suppose that $g(\by|\bx,\sigma)$ is an $m$-dimensional multivariate Gaussian proposal distribution, $\by \sim MVN(\bx, \, \sigma^2 \bA)$, where $\bA $ does not depend on $\sigma$. Then a lower bound on $c^*/\sigma^*$ is $(mp^*)^{-1}$.
\end{prop}

\noindent {\bf Proof:} Differentiation of $g(\by|\, \bx,\sigma) = (2 \pi \sigma^2 )^{-m/2} | \bA | ^{-1/2} \exp \{ - {\frac{1}{2} } (\by - \bx )' \bA ^{-1} (\by - \bx ) / \sigma ^2 \}  $ gives $dg(\by|\, \bx,\sigma)/  d\sigma = \{ \sigma ^{-3} (\by - \bx )' \bA ^{-1} (\by - \bx ) \, - \, m\sigma ^{-1} \} g(\by|\, \bx,\sigma)$.
Substituting this expression in Equation (\ref{dpds}) gives
\begin{equation}\label{eqn:diff}
dp(\sigma)/d\sigma = -mp(\sigma)/\sigma + \phi,
\end{equation}
 where
\begin{equation}\label{eqn:phi}
 \phi =  \sigma ^{-3} \int \int \mbox{min}\left\{ \frac{f^{\#}(\by)}{f^{\#}(\bx) },\,1 \right\} \,  (\by - \bx )' \bA ^{-1} (\by - \bx ) \, g(\by \, |\, \bx,\sigma) \, f(\bx) \, d\by \, d\bx.
\end{equation}
Since $\phi>0 $ is positive,
$dp(\sigma)/d\sigma > -mp(\sigma)/\sigma.$
It follows that $c^*/\sigma^* > (mp^*)^{-1}$, as $c^* = -1/ [dp(\sigma)/d\sigma ] _{\sigma =\sigma^*}$ and $p(\sigma^*)=p^*$. \, \, \, \, \hfill $\Box $

\begin{prop}
\label{PropP0}Let $m=1$ and suppose the conditions of Proposition \ref{PropBound} hold and that $f(\cdot)$ has finite variance. Then $c^*/\sigma^* \rightarrow (p^*)^{-1}$ as $\sigma^* \rightarrow \infty$, where $(\sigma^*)^2 \bA$ is the variance of the proposal distribution.
\end{prop}

\noindent {\bf Proof:} In appendix.

\begin{prop}
\label{PropP1} Suppose that the proposal distribution is the multivariate Gaussian distribution defined in Proposition \ref{PropBound} and the target distribution $f(\bx)$ is continuous over the support of $\bx$. If $\sigma \rightarrow 0$ as $p(\sigma ) \rightarrow 1$, then  $c^*/\sigma^* \approx 1/(1-p^*)$ as $p^* \rightarrow 1$. 
\end{prop}

\noindent {\bf Proof:} In appendix.
\\

We are required to adopt some relationship between $c^*/\sigma^*$ and $p^*$ that will be taken as representative of the relationship for target distributions in general. Under mild regularity conditions on the target distribution, $p^* \rightarrow 0$ only as $\sigma^* \rightarrow \infty$. Hence, when $m=1$, Proposition \ref{PropP0} implies that the relationship should satisfy $c^*/\sigma^* \rightarrow 1/p^*$ as $p^* \rightarrow 0$. Also, from Propositions \ref{PropBound} and \ref{PropP1},  $c^*/\sigma^*$ should exceed $1/(mp^*)$ for all $p^*$ and $c^*/\sigma^* \rightarrow 1/(1-p^*)$ as $p^* \rightarrow 1$. Thus, when the target distribution is a univariate distribution, so that $m=1$, a natural relationship to consider is
\begin{equation}\label{eqn:uni}
c^*/\sigma ^* \approx \frac{1}{p^*(1-p^*)},
\end{equation}
as this is the simplest function that meets these conditions.

We examined the relationship between $c^* / \sigma ^*$ and $p^*$ for a broad range of univariate target distributions, based on a univariate Gaussian random walk proposal with variance $\sigma^2$. Specifically these distributions were the standard normal, t (with 5 d.f.), Cauchy, uniform, logistic and double exponential ($f(x)=0.5e^{-|x|}$, $-\infty <x<\infty$) distributions, a Gamma(5, 1) and Beta(3, 7) distribution,
and a mixture of two normal distributions: $\frac{1}{2} N(0,1) + \frac{1}{2} N(5,5)$.
For each target distribution, and for a range of values of $p^*\in[0.05, 0.95]$,
Monte Carlo methods were used to determine $c^*/\sigma^*$ by first solving Equation (\ref{eqn:p}) for $\sigma$ and then evaluating $dp/d\sigma$ at that value of $\sigma$, using Equation (\ref{dpds}).  Large sample sizes were used to ensure that Monte Carlo variability was negligible.

\begin{figure}
\begin{center}
\includegraphics[clip = true , viewport=0 30 600 420, scale=.4]{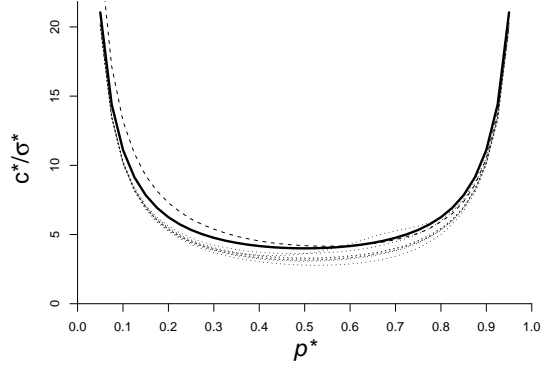}
\caption{Plots of $c^* / \sigma ^*$ against $p^*$ for nine univariate distributions: (dotted lines) standard normal, t (with 5 d.f.),  uniform, logistic and double exponential distributions, a Gamma(5, 1) and Beta(3, 7) distribution, a normal mixture $\frac{1}{2} N(0,1) + \frac{1}{2} N(5,5)$, and (dashed line) a standard Cauchy. Solid line denotes the relationship $c^*/\sigma^*=1/[p^*(1-p*)]$.}\label{figure1}
\end{center}
\end{figure}

Figure \ref{figure1} plots $c^* / \sigma ^*$ against $p^*$ (dotted lines) for each of these distributions.
The closeness of the nine lines indicates that the relationship is broadly similar across these distributions, although the values of $c^*/\sigma^*$ for the Cauchy distribution (dashed line) are slightly larger than the others for low $p^*$.
The solid line in Figure \ref{figure1} is the relationship between $c^* / \sigma ^*$ and $p^*$ given by equation (\ref{eqn:uni}). The figure indicates that (\ref{eqn:uni}) is a good choice for this relationship, although many other choices would also be satisfactory and could be made without greatly affecting the performance of the method. A useful feature of  (\ref{eqn:uni}) is that it generally yields an estimate of $c^*$ that is a little too large rather than too small: it is better to overestimate the steplength constant  than to underestimate it.

Returning to multivariate distributions, \shortciteN{roberts+gg97} considered $m$-dimensional target distributions of the form
\begin{equation}\label{eqn:prodind}
f(\bx) = h(x_1)\/h(x_2)\ldots h(x_m)
\end{equation}
for some one-dimensional smooth density $h$, where $\bx=(x_1,\ldots,x_m)'$. They showed that if the proposal distribution is an $m$-dimensional Gaussian distribution, $\by \sim MVN(\bx , \sigma^2 \bI_m )$, then $p(\sigma) = 2\Phi (-\sigma B m^{1/2}/2)$ as $m \rightarrow \infty$, where $B>0$  is a constant that depends on $h$. \shortciteN{roberts+r01} derive similar results for the case where the target distribution is a multivariate Gaussian and the proposal distribution is an $m$-dimensional multivariate normal distribution, $\by \sim MVN(\bx, \, \sigma^2 \bA)$, or if $f(\bx) = \prod _{i=1} ^m C_i h(C_i x_i )$, where the $ \{ C_i \} $ are i.i.d draws from some fixed distribution. They show that as $m \rightarrow \infty $, the value of $p(\sigma)$ tends to  $2\Phi (-\beta \sigma )$ for some positive constant $\beta$. The following proposition gives  $c^*/\sigma^*$ whenever $p(\sigma)$ has this form.

\begin{prop}
\label{PropMult} Suppose that  $p(\sigma) = 2 \Phi (-\beta \sigma)$, where $\beta>0$ is a constant and $\Phi $ is the cdf of the standard normal distribution. Then
\begin{equation}
c^* /\sigma ^* = (2\pi)^{1/2}\exp (\alpha ^2 /2)/(2 \alpha),
\end{equation}
where $\alpha = -\Phi ^{-1} (p^* /2)$.
\end{prop}

\noindent {\bf Proof:} Differentiating $p(\sigma) =2 \int _{-\infty}^{-\beta \sigma } (2\pi)^{-1/2}\exp (-z ^2 /2)\/dz$ gives $dp(\sigma) /d\sigma = -2\beta (2\pi)^{-1/2}$ $\exp \{ -(\beta \sigma ) ^2 /2)$. Write $\alpha = \beta \sigma ^*$, so that  $\alpha = -\Phi ^{-1} (p^* /2)$. The proposition  follows as $c^* / \sigma ^* = -1/ \{ \sigma ^* \/ [ dp(\sigma) / d\sigma ] _{\sigma =\sigma^*} \}$. \, \, \hfill $\Box $\\

To gain an impression of how $c^*/\sigma ^*$ varies with the dimension, $m$, for multivariate target distributions,
we consider both target distributions of the form (\ref{eqn:prodind}), and $m$-dimensional multivariate-$t$ distributions on $\nu$ degrees of freedom.
For fixed $p^*$, and with a multivariate Gaussian random-walk proposal distribution with covariance matrix $\sigma^2 \bI_m $, experimentation indicated that $c^*/\sigma ^*$ was close to a linear function of $1/m^*$. Here, $m^*=m$ for target distributions of the form (\ref{eqn:prodind}), and $m^*=\min\{m,\nu\}$ for the multivariate-$t$ target distributions.
If we assume that Proposition \ref{PropMult} holds as $m^* \rightarrow \infty$, and that Equation (\ref{eqn:uni}) also holds when $m^*=1$, then this determines the linear function as
\begin{equation}\label{eqn:mix}
c^*/\sigma ^* \, = \,  \left( 1- \frac{1}{m^*} \right)  \frac{ (2\pi )^{1/2} e ^{\alpha ^2 /2 }}{2\alpha} \, + \, \frac{1}{m^*p^*(1-p^*)},
\end{equation}
where $\alpha = -\Phi ^{-1} (p^* /2)$.\vspace{.2in}

\begin{figure}
\begin{center}
\includegraphics[width=8cm]{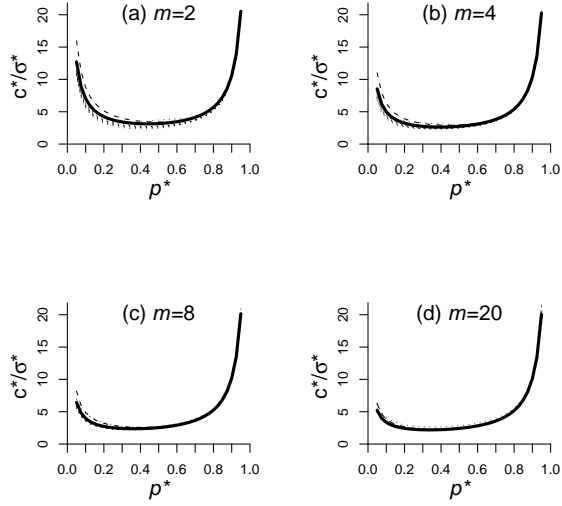} \\
\caption{Plots of $c^* / \sigma ^*$ against $p^*$ for $m$-dimensional multivariate distributions of the form $f(\bx)=\prod_{i=1}^mh(x_i)$ where $h(\cdot)$ is given by the nine univariate distributions in Figure \ref{figure1} (dotted lines). Dashed line denotes the standard Cauchy distribution. Solid line denotes the relationship given by  (\ref{eqn:mix}). Panels (a)--(d) represent $m=2, 4, 8$ and $20$ dimensions.}\label{figure2}
\end{center}
\end{figure}

To examine the usefulness of (\ref{eqn:mix}) for target distributions of the form $f(\bx)=\prod_{i=1}^mh(x_i)$, each of the univariate distributions considered in Figure \ref{figure1} was taken in turn as the component distribution $h(\cdot)$.
As before, Monte Carlo methods were used to determine $c^*/\sigma ^*$ for a range of values of $p^*\in[0.05 , 0.95]$, and the resulting relationships are shown in
Figure \ref{figure2}
for $m=2$, $4$, $8$ and $20$ dimensions. In each panel (a)--(d) the highest (dashed) line corresponds to the Cauchy distribution and the dotted lines to the other target distributions. For each model dimension, the relationship between $c^*/\sigma^*$ and $p^*$ is similar for all distributions. The solid line illustrates the linear function (\ref{eqn:mix}), which exhibits a strong similarity with the other curves, implying it models the relationship across model dimensions well for these models. As with the univariate case (Figure \ref{figure1}), the form of (\ref{eqn:mix}) generally represents a practically convenient overestimate of $c^*$.

\begin{figure}
\begin{center}
\includegraphics[clip = true , viewport=0 240 420 420, scale=.6]{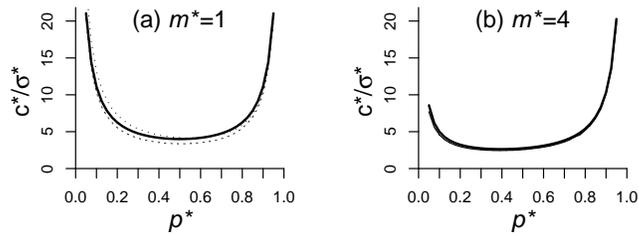}
\caption{Plots of $c^* / \sigma ^*$ against $p^*$ for univariate and multivariate $t$ distributions with $\nu$ degrees of freedom. Panel (a) shows $(m=1,\nu =1)$, $(m=4,\nu =1)$ and $(m=1,\nu =4)$, so that  $m^* =1$. Panel (b) shows $(m=4,\nu =4)$, $(m=4,\nu =8)$ and $(m=8,\nu =4)$, so that $m^* =4$.
Solid line denotes the relationship given by  (\ref{eqn:mix}). }\label{figure3}
\end{center}
\end{figure}

To examine the usefulness of  (\ref{eqn:mix}) for multivariate $t$ target distributions, similar calculations were performed for various combinations of model dimension, $m$, and degrees of freedom, $\nu$.
The results are presented in Figure \ref{figure3} for $m^*=1$ (panel a) and (b) $m^*=4$ (panel b), where in each case the solid line  illustrates the relationship given by (\ref{eqn:mix}).
In Figure \ref{figure3}(a) the top curve is from the Cauchy distribution $(m =1, \nu=1)$ and the bottom curve is actually {\em two} near-identical curves given by $(m=4,\nu=1)$ and $(m=1,\nu=4)$. In Figure \ref{figure3}(b) all curves are visually indistinguishable.
Hence, using $m^*=\min\{m,\nu\}$ appears sensible for $t$ distributions, and, as such, provides strong support for (\ref{eqn:mix}) as a representation of the relationship between $c^*/\sigma^*$ and $p^*$.

In order to use Equation (\ref{eqn:mix}), the value of $m^*$ must be specified.
However, consideration of multivariate $t$-distributions with few degrees of freedom indicates that common choices of $p^*$ ($p^*=0.44$ for a univariate target and $p^*=0.234$ for a multivariate target of moderate or large dimension)
 are not always appropriate. For example, with a multivariate Cauchy target distribution, $f^{\#}(\by)/f^{\#}(\bx)$ is the ratio of two multivariate normal distributions multiplied by the ratio of two univariate normal distributions. Choosing $p^*$ on the basis of the dimension of the multivariate distributions is a poor approach as this ignores the latter ratio. The considerations necessary for choosing $p^*$ should also give a suitable choice of $m^*$.

If $\widehat{\sigma}_c$ is the estimate of $\sigma^*$ after $i$ steps of a Robbins-Monro search with a steplength constant of $c$, then var$(\widehat{\sigma}_c) = p^*(1-p^*)c^2/ \{ i (2c/c^* -1) \} $ (\shortciteNP{hodgesl55}). This is minimized when $c=c^*$ so the efficiency of a search is defined to be
\begin{equation}\label{eqn:eff}
\frac{\mbox{var}(\widehat{\sigma}_{c^*})}{\mbox{var}(\widehat{\sigma}_c)}\, \times \, 100 \% \, = \, \frac{(2c-c^*)c^*}{c^2} \, \times \, 100 \% .
\end{equation}
Efficiency declines slowly if $c^*$ is overestimated: efficiency is 75\% when $c=2c^*$
and $c$ has to exceed more than $3.4c^*$ before efficency drops below 50\%. As such, if a suitable choice for $m^*$ is unclear, setting $m^*$ to a small value (such as $m^*=1$) should not be a disaster as $c^*$ will be overestimated.  When the choice of $m^*$ is unclear, the length of the Markov chain burn-in phase
might be doubled, which would more than compensate for a drop in efficiency of up to 50\%.

In summary,
Figures \ref{figure1}--\ref{figure3} show that $c^*/\sigma^*$ is largely determined by the values of $p^*$ and $m^*$ for a range of distributions, and Propositions 1--4 suggest that this should hold more generally. Equation (\ref{eqn:mix}) gives a good estimate of $c^*/\sigma^*$ and it will be used in our implementation of the Robbins-Monro process within an MCMC sampler. That is, when $\sigma$ is the estimate of $\sigma^*$ that the search currently gives, we will put
\begin{equation}\label{eqn:steplength}
c \, = \,  \sigma \left\{\left( 1- \frac{1}{m^*} \right) . \frac{ (2\pi )^{1/2} e ^{\alpha ^2 /2 }}{2\alpha} \, + \, \frac{1}{m^*p^*(1-p^*)} \right\} ,
\end{equation}
as the steplength constant for the next step of the search.

When the steplength constant is given by (\ref{eqn:steplength}), calculation shows that the efficiency of the search for each of the distributions considered in Figure \ref{figure1} is at least 96\% when $p^* =0.234$ and at least 91\% when $p^* =0.440$. In the context of forming confidence intervals, Garthwaite and Buckland (1992) found the efficiency of the Robbins-Monro was generally noticeably lower (but adequate), typically around 70-80\% for 95\% confidence intervals. The improved efficiency here is because the proposal distribution is fixed as a multivariate normal distribution and this largely determines the optimal value of the steplength constant.\vspace{.2in}

\section{Search algorithms for adaptive optimal scaling}
\label{sec:example}

In this section we describe the implementation of the Robbins-Monro search process within MCMC algorithms.
We initially consider univariate and multivariate target distributions where all model parameters are updated simultaneously using Metropolis-Hastings updates.
We then consider a multivariate target within the Metropolis-Hastings within Gibbs sampler framework.
The performances of the algorithms are examined via examples in Section 5.

\subsection{Univariate target distributions}
\label{sec:searchuni}

For the univariate target distribution $f(x) \propto f^{\#} (x)$, under a Gaussian RWMH sampler with proposal distribution  $y\sim N(x, \sigma^2)$, we aim to use the Robbins-Monro process to determine the value of $\sigma=\sigma^*$ so that the OAP of the sampler is $p^*$. Following \citeN{roberts+r01} we suppose that $p^*=0.44$ is appropriate.

The basic strategy is to continually improve the estimate of $\sigma^*$ at each step of the Markov chain.
If $\sigma _i$ denotes the estimate of $\sigma^*$ after the $i$-th step of the search, then we set
\begin{equation}\label{eqn:RMrule}
\sigma _{i+1} = \left\{ \begin{array}{ll}
\sigma _i + c(1-p^*)/i & \mbox{ if $y$ is accepted}  \\
\sigma _i - cp^*/i & \mbox{ if $y$ is rejected}
\end{array} \right.
\end{equation}
where $c =  \sigma _i / \{ p^* (1-p^*) \} $ (c.f. equation (\ref{eqn:uni})). 
Starting values for a search can be arbitrary (e.g. $\sigma_1 = 1$) or more considered, such as
an estimated standard deviation of $f(\cdot)$.
Either way,
$\sigma_1$ need not be well-chosen, as
the Robbins-Monro process can be monitored and a search restarted if the starting value seems poor (e.g. \shortciteNP{garthwaite96}, \shortciteNP{matsuio99}).
Otherwise $\sigma_i\rightarrow\sigma^*$ can take a long time to converge as the step size decreases with $i$.

On a restart, the most recent estimate of $\sigma^*$ is taken as the starting value and the value of $i$ is reset.
Note that we start (and restart) a search with $i=n_0$, where $n_0$ is a moderate size so as
to avoid too rapid steplength changes in the early stages of a search  (e.g. steps would halve in size between $i=1$ and $i=2$).
We choose $n_0$ to be the integer closest to $5/ \{ p^* (1-p^*) \} $, which works well in practice.
We also choose to restart the search if the estimate of $\sigma^*$ changes by a factor of 3
from its value from when the search started (or last restarted). Many other criteria for restarts would also work well, as the only requirement is to restart if a poor starting value has been used.

It seems conceivable that a search might oscillate between $\sigma$ tripling in value and reducing in value by two-thirds, so that the search is continually restarting (though this has never happened in our experience). To ensure that this cannot happen, our algorithm records the number of restarts resulting from $\sigma$ tripling and the number resulting from it reducing by two-thirds. Should both these numbers reach 5, then the process is not restarted again. We also do not restart if more than 100 steps have been taken since the last restart, as taking 100 steps without restarting suggests a reasonable starting point has been used. These decision rules are obviously arbitrary to a degree, but they work well in practice in that they seldom need to be enforced. At the same time, the rules mean that the size of steps definitely shrink to zero, so the procedure satisfies the diminishing adaptation criterion.

\subsection{Multivariate Metropolis-Hastings updates}
\label{sec:searchfull}
We now consider a multivariate target distribution, $f(\bx) \propto f^{\#} (\bx)$, where all components of $\bx$ are updated simultaneously using a Gaussian random walk proposal  $MVN(\bx, \, \sigma^2 \bA )$, for some positive-definite matrix $\bA$, where
the purpose of the Robbins-Monro search is to find a value of $\sigma$ that gives an OAP of $p^*$.
Following \shortciteN{roberts+gg97} we set $p^*=0.234$.
The search procedure is fundamentally the same as for the univariate target distribution. A starting value for $\sigma $ is again required and an arbitrary value may be chosen, the value for $n_0$ is determined in the same way as before, and the same method of monitoring and restarting searches is followed. Other details differ slightly. The steplength constant is determined by Equation (\ref{eqn:steplength}), so
\begin{equation}
c \, = \,  \sigma _i \left\{\left( 1- \frac{1}{m^*} \right) . \frac{ (2\pi )^{1/2} e ^{\alpha ^2 /2 }}{2\alpha} \, + \, \frac{1}{m^*p^*(1-p^*)} \right\} \vspace{.2in}
\end{equation}
where $\alpha=-\Phi^{-1}(p^*/2)$.
Usually, the value of $m^*$ should equal $m$, the dimension of $\bx$. However, if $j$ components of $\bx$ are related by some variate (e.g. a variance in the case of $t$-distributions) that has $\nu$ degrees of freedom, and $\nu <j$, then we suggest reducing $m^*$ by $j-\nu$. This is in line with the results illustrated in Figure \ref{figure3}. If there is doubt as to the appropriate value of $m^*$, the analysis in Section \ref{sec:est} advocates setting $m^*$ too small rather than too large.

Turning to the choice of $\bA$, in many circumstances the convergence rate of the Markov chain can only be optimized if $\bA$ is proportional to the covariance matrix of $\bx$. This covariance matrix, $\bSig $ say, is typically unknown, but its value may be estimated as the Markov chain runs (\shortciteNP{craiu+ry09}). After each iteration of the chain, $\bA$ may be set equal to the current estimate of $\bSig$.
Thus we can set
\begin{equation}\label{eqn:sig}
\widehat{\bSig}_i = \left\{
\begin{array}{ll}
\bI_m, &i\leq 100\\
\frac{1}{i-1}\sum_{j=1}^i (\bx_j-\bar{\bx}_i)(\bx_j-\bar{\bx}_i)' &i > 100.
\end{array}\right.
\end{equation}
where the computation for updating $\hat{\Sigma}_i$ is reduced by using the recursions
$$
\bar{\bx}_i = \frac{1}{i}\left[(i-1)\bar{\bx}_{i-1}+\bx_i\right]
$$
and
$$
\widehat{\bSig}_i = \frac{i-2}{i-1}\widehat{\bSig}_{i-1}+\bar{\bx}_{i-1}\bar{\bx}'_{i-1}-\frac{i}{i-1}\bar{\bx}_{i}\bar{\bx}'_{i}+
\frac{1}{i-1}\bx_i\bx'_i . $$
\shortciteN{haario+st01} suggest $\widehat{\bSig}_i+\epsilon \bI_m $  (where $\epsilon >0$) as a positive-definite estimate of $\bSig$.
In our implementation we follow this approach with $\epsilon=\sigma ^2_i /i $,  so that $\bA_i = \widehat{\bSig}_i +\sigma ^2_i  \bI_m /i$ as the estimate of $\bSig$. Thus,  $MVN(\bx ,\, \sigma ^2_i  \bA _i )$ is used as the proposal distribution after $i$ steps of the Robbins-Monro search.

When the dimension of $\bx$ is large,
a substantial number of iterations of the Markov chain may be needed before the estimate of $\bSig $ stabilizes. \citeN{roberts+r09} give examples where 400,000 iterations were needed when $m=100$, and nearly 2 million were needed when $m=200$. For the Robbins-Monro process to converge to the correct value of $\sigma ^*$, it must not converge before the estimate of $\bSig$ stabilizes. Similarly, many sampler iterations are typically needed to effectively explore the parameter space when there are many parameters, and proposal acceptance probabilities may vary dramatically over this space. Hence, even after the estimate of $\bSig$ is fairly stable, the estimate of $\sigma$ should converge slowly if it is to reflect the overall optimum for the parameter space.

To achieve this, we propose that the magnitude of steps should not be reduced below some pre-fixed limit until the estimate of $\bSig$ is reasonably stable, and after that the stepsize  should be reduced slowly. In our algorithm, for $i>200$ we set
\begin{equation}\label{eqn:cont}
\sigma _{i+1} = \left\{ \begin{array}{ll}
\sigma _i + c(1-p^*)/{\mbox{max}}\{200, i/m\} & \mbox{ if $\by$ is accepted}  \\
\sigma _i - cp^*/{\mbox{max}}\{200, i/m\} & \mbox{ if $\by$ is rejected.}
\end{array} \right.
\end{equation}
which works well in practice. There are many alternative choices that would also be satisfactory. Monitoring of the traceplots for the parameters should be carried out to ensure
that estimates of $\bSig$ have stabilized.

\subsection{Metropolis-Hastings within Gibbs}
\label{sec:searchMG}

In Metropolis-Hastings within Gibbs, $\bx$ is partitioned into components, $\bx' =(\bx'_1, \ldots, \bx_j')$, and each component is sampled sequentially (conditional on other components), using Gibbs sampler component updates where possible, but otherwise using Metropolis-Hastings updates. In the latter case, we suppose the proposal distribution is of the form $N(x_k,\,\sigma^2)$ for scalar components, or $MVN(\bx_k, \, \sigma^2_k \bA_k)$ for vectors, $k=1,\ldots,j$. The appropriate value of $\sigma ^2$ varies from component to component, so if there are $q\leq j$ components that require a Metropolis-Hastings sampler, then $q$ Robbins-Monro searches are conducted while the sampler runs, each one searching for $\sigma^* $ for one particular component. The searches are conducted using the methods given in Sections 4.1 and 4.2.

\section{Examples}
\label{sec:examples}

\subsection{Univariate target distributions}
\label{sec:exUni}

The search algorithm given in Section \ref{sec:searchuni} was applied in  turn to each of the nine univariate target distributions considered earlier (c.f. Figure \ref{figure1}).
Two hundred samplers were run for each target distribution, and a search for the value of $\sigma ^*$ that gave an OAP of 0.44 was conducted within each chain. The starting point for each search was obtained by randomly setting $\sigma_1\sim\mbox{Exp}(1)$.
As a reasonable aim of the search process is that it should give a good estimate of $\sigma ^*$ within 2000 steps, each chain was run for 2000 iterations. The final estimate of $\sigma ^*$ and the acceptance rate of the sampler over the last 1000 iterations was recorded for each chain. The results are summarized in Table \ref{table3}.

\begin{table}[h]
\caption{ Sampler performance based on 200 chain replicates of length 2000 iterations, under a target acceptance probability of $p^*=0.44$, for each of the nine univariate target distributions in Figure \ref{figure1}.
Optimal (true) values of $\sigma^*$, and 0.05, 0.5 and 0.95 quantiles of the empirical distribution of the estimates of $\sigma^*$ are given, together with quantiles of the sampler acceptance rates in the last 1000 iterations.}\label{table3}
\begin{center}
\begin{tabular}{|lccrrrrcrrrc|}\hline
{\em Target} & & Optimal & & \multicolumn{3}{c}{$\sigma^*$ quantile} & & \multicolumn{3}{c}{OAP quantile} & \\
\cline{5-7} \cline{9-11}
{\em distribution} & & $\sigma^*$ & &\multicolumn{1}{c}{0.05} & \multicolumn{1}{c}{median} & \multicolumn{1}{c}{0.95} & & \multicolumn{1}{c}{0.05} & \multicolumn{1}{c}{median} & \multicolumn{1}{c}{0.95} & \\
\hline
$N(0, 1 )$           & &  2.42 & & 2.31 & 2.43 & 2.56 & & 0.417 & 0.443 & 0.468 & \\
$t$-dist. (5 d.f.)   & &  2.71 & & 2.54 & 2.73 & 2.89 & & 0.413 & 0.441 & 0.470 & \\
Cauchy               & &  4.39 & & 3.69 & 4.25 & 5.03 & & 0.389 & 0.443 & 0.501 & \\
Logistic             & &  4.05 & & 3.82 & 4.05 & 4.33 & & 0.417 & 0.442 & 0.467 & \\
Double exponential   & &  2.70 & & 2.52 & 2.70 & 2.93 & & 0.413 & 0.439 & 0.465 & \\
Gamma(5,1)           & &  4.98 & & 4.62 & 4.96 & 5.28 & & 0.414 & 0.443 & 0.467 & \\
Beta(3,7)        & &  0.335 & & 0.311 & 0.335 & 0.355 & & 0.417 & 0.440 & 0.466 & \\
Uniform          & &  0.806 & & 0.764 & 0.807 & 0.849 & & 0.418 & 0.442 & 0.464 & \\
$\frac{1}{2} N(0,1) + \frac{1}{2} N(5,5)$ & & 6.07 & & 5.59 & 6.10 & 6.50 &  & 0.415 & 0.442 & 0.468 & \\[1mm]
\hline
\end{tabular}
\end{center}
\end{table}

For each of the nine univariate distributions,
the second column in Table \ref{table3} provides the theoretical value of $\sigma ^*$ and the next three columns give the 0.05, 0.50 and 0.95 quantiles of the final estimate of $\sigma ^*$ from each search. The last three columns present the same quantiles of the acceptance rates in the last 1000 steps of each search. The results indicate that the Robbins-Monro search has low bias and good accuracy: the median estimate of $\sigma ^*$ is close to $\sigma ^*$ for each of the nine distributions, the median values of the OAP are close to their target of 0.44, and the 0.05 and 0.95 quantiles for both $\sigma ^*$ and the OAP are quite close together. This performance clearly exceeds requirements, as the efficiency of the Metropolis Hastings algorithm is not sensitive to the precise value of $p^*$.

\subsection{Multivariate Metropolis-Hastings updates}
\label{sec:exFull}

We follow the example of \citeN{roberts+r09} in which the target distribution is
$f(\bx) = MVN(0, \bSig )$, where $\bSig =\mbox{\bf M}\mbox{\bf M}'$ and $\mbox{\bf M}$ is an $m \times m$ matrix whose elements are generated randomly from a $N(0,1)$ distribution.
This target distribution is somewhat pathological, in that typically $\bSig$ will be close to singular, and so
we refer to this as the {\em ill-conditioned case}.
We also consider a modified target distribution, in which each diagonal element of $\bSig$ is increased by 1\%, and refer to this as the {\em  better-conditioned case}. It is perhaps more representative of target distributions that arise in practice.
Sampling from these target distributions is difficult when the dimension $m$ is moderate to large, especially with the ill-conditioned case. Here we set $m=50$.

For each case we consider three versions of the RWMH
sampler. The first version (the {\em RM method}) is
the algorithm proposed in Section 4.2, where the proposal distribution has the form $MVN(\bx_{i-1}, \sigma_i^2\bA_i)$, where $\sigma_i$ is estimated using the Robbins-Monro procedure and $\bA_i=\widehat{\bSig}_i+\sigma_i^2\bI_m/i $, with $\widehat{\bSig}_i$ empirically estimated from Equation (\ref{eqn:sig}).
In theory, the optimal proposal distribution is $MVN(\bx_{i-1}, 2.38^2 \bSig/m )$ \shortcite{roberts+gg97}. The second sampler version (the {\em Optimal method}) implements a RWMH sampler with this (fixed) proposal distribution.
This sampler is unavailable in practice, as the true value of $\bSig$ is typically unknown, but its performance provides a benchmark for the other methods.
The final sampler version (the {\em fixed-scaling} method) implements the RWMH sampler with covariance matrix $2.38^2\bA_i/m$, where $\bA_i$ is estimated as in the RM method. This sampler is part way between the RM and the Optimal methods.

Ten replicate samplers, each of length 100,000 iterations,
were run for both the ill-conditioned and the better-conditioned case, and for each of the sampler variants. Results are based on discarding the first half of a chain as burn-in.
For the Robbins-Monro algorithm, we set the desired overall acceptance probability (OAP) equal to $p^*=0.234$, $n_0 =20$ and $m^* =50$.
As a measure of algorithm efficiency, we follow \citeN{roberts+r09}
in monitoring the
integrated auto-correlation time (ACT)
(e.g. \citeNP{roberts+r01}).
We also monitor the average squared jumping distance  (ASD) between the iterates of the chain. A smaller value of the ACT indicates less auto-correlation, and hence greater efficiency. Similarly, the larger the jumping distance, the faster the mixing of the chain. All ACT and ASD values are
calculated using full length of the chain (including the burn-in period).

A summary of the results is provided in
Table \ref{table2}, 
which includes specific results for the first coordinate, $x_1$. With the better-conditioned case, results for all three methods are highly satisfactory. The RM method consistently estimates $\sigma^2$ with good accuracy and the OAPs for all methods are close to the target value of 0.234.  The Optimal method benefits from the unrealistic advantage of knowing $\Sigma$, and it has a noticeably better ASD than the other methods. However, all three methods give good estimates of the mean and standard deviation of $x_1$ (whose true values are 0 and 7.48 respectively).   Results for the ill-conditioned case are more diverse. The estimates of $\Sigma$ are poor by construction, which hinders the RM method and the fixed-scaling method, both of which must use these estimates. The RM method compensates by setting $\sigma^2$ to a low value and this enables the method to attain the target acceptance probability of 0.234. In contrast the fixed-scaling method
achieved an OAP of only 0.14.

\begin{table}
\caption{ Summaries of RWMH sampler performance under three sampler variants  using a multivariate
normal proposal distribution,  for both better- and ill-conditioned cases. Parentheses indicate Monte Carlo standard errors based on 10 sampler replicates.
 Columns correspond to
the mean value of $\sigma^2$; the overall acceptance probability (OAP); posterior mean and standard deviation for the component $x_1$,
the integrated autocorrelation times (ACT) for the parameter $x_1$; and the average squared distances (ASD) between the iterates of the parameter $x_1$.}\label{table2}
\begin{center}
\begin{tabular}{|lcccccc|}\hline
  & & & \multicolumn{4}{c|}{Statistics for $x_1$} \\ \cline{4-7}
  Version  &$\sigma^2$ &OAP & mean   & sd  &ACT &ASD\\
  \hline
  \multicolumn{4}{|l}{\em Better-conditioned case} & & & \\
  RM method & 0.126 (0.01) & 0.233 (0.001) & \phantom{-}0.14 (0.33) & 7.09 (0.18) &77.98 (1.08)&1.12 (0.09)\\
  Optimal $\bSig$ & 0.113 (\phantom{1}--\phantom{1}) & 0.239 (0.001) &\phantom{-}0.09 (0.30) & 7.60 (0.17) & 75.08 (0.83)& 1.48 (0.01)\\
  Fixed-scaling & 0.113 (\phantom{1}--\phantom{1}) & 0.258 (0.001) &-0.01  (0.44) & 7.01 (0.22) & 78.29 (1.03)&1.08 (0.09) \\&&&&&&  \\
  \multicolumn{4}{|l}{\em Ill-conditioned case} & & & \\
  RM method & 0.070 (0.001) & 0.233 (0.006) & \phantom{-}0.12 (0.52) & 6.63 (0.28)  & 89.13 (0.93)&0.47 (0.04)\\
  Optimal $\bSig$ &0.113 (\phantom{1-}--\phantom{-1}) &0.239 (0.001) & -0.02 (0.34) &7.41 (0.14) &74.20 (0.93) &1.45 (0.02) \\
  Fixed-scaling &0.113 (\phantom{1-}--\phantom{-1}) &0.141 (0.001) & -0.14 (0.48) &6.37 (0.33) &89.05 (0.56) &0.44 (0.03) \\
  \hline
\end{tabular}
\end{center}
\end{table}

\begin{figure}[t]
\begin{center}
\psfrag{RM sigmai}[l][l]{}
\psfrag{iterations}[l][l]{{\tiny Iterations }}
\begin{tabular}{ll}
\psfrag{x}[t][t]{{\tiny $\sigma^2$}}
\includegraphics[width=7cm, height=5cm]{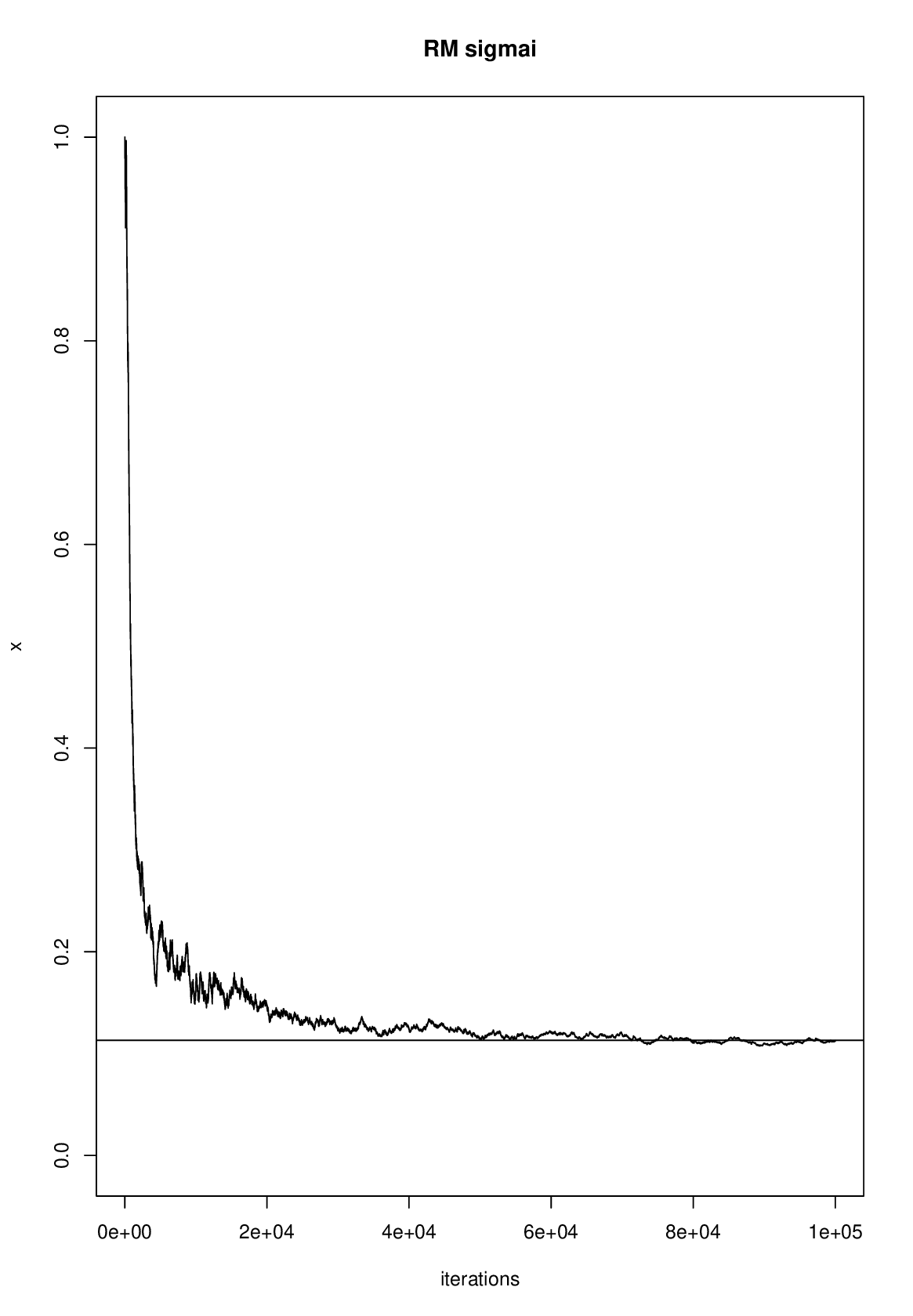} &\psfrag{x}[t][t]{{\tiny $\sigma^2$}} \includegraphics[width=7cm, height=5cm]{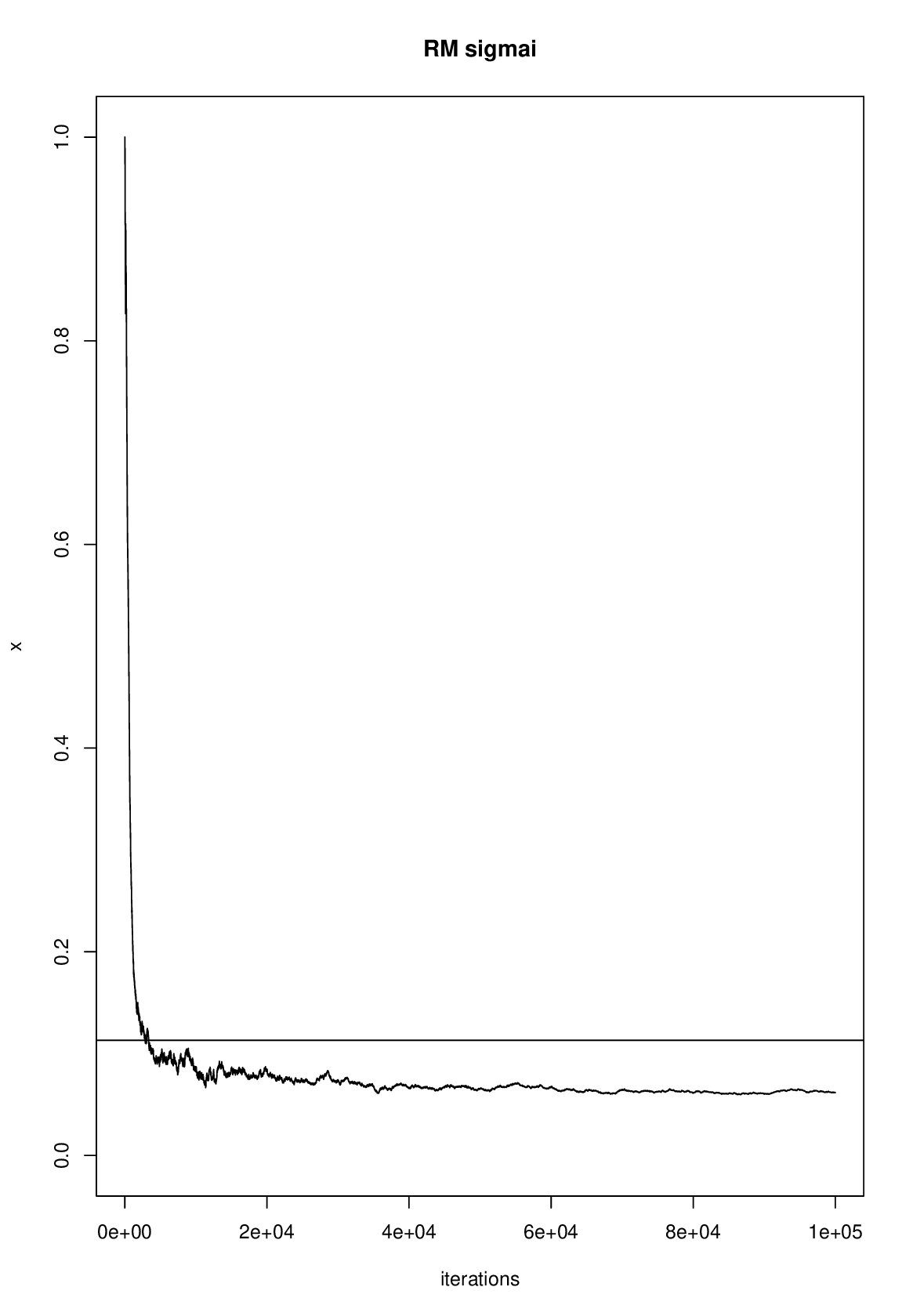} \\
\psfrag{x}[t][t]{{\tiny OAP}}
\includegraphics[width=7cm, height=5cm]{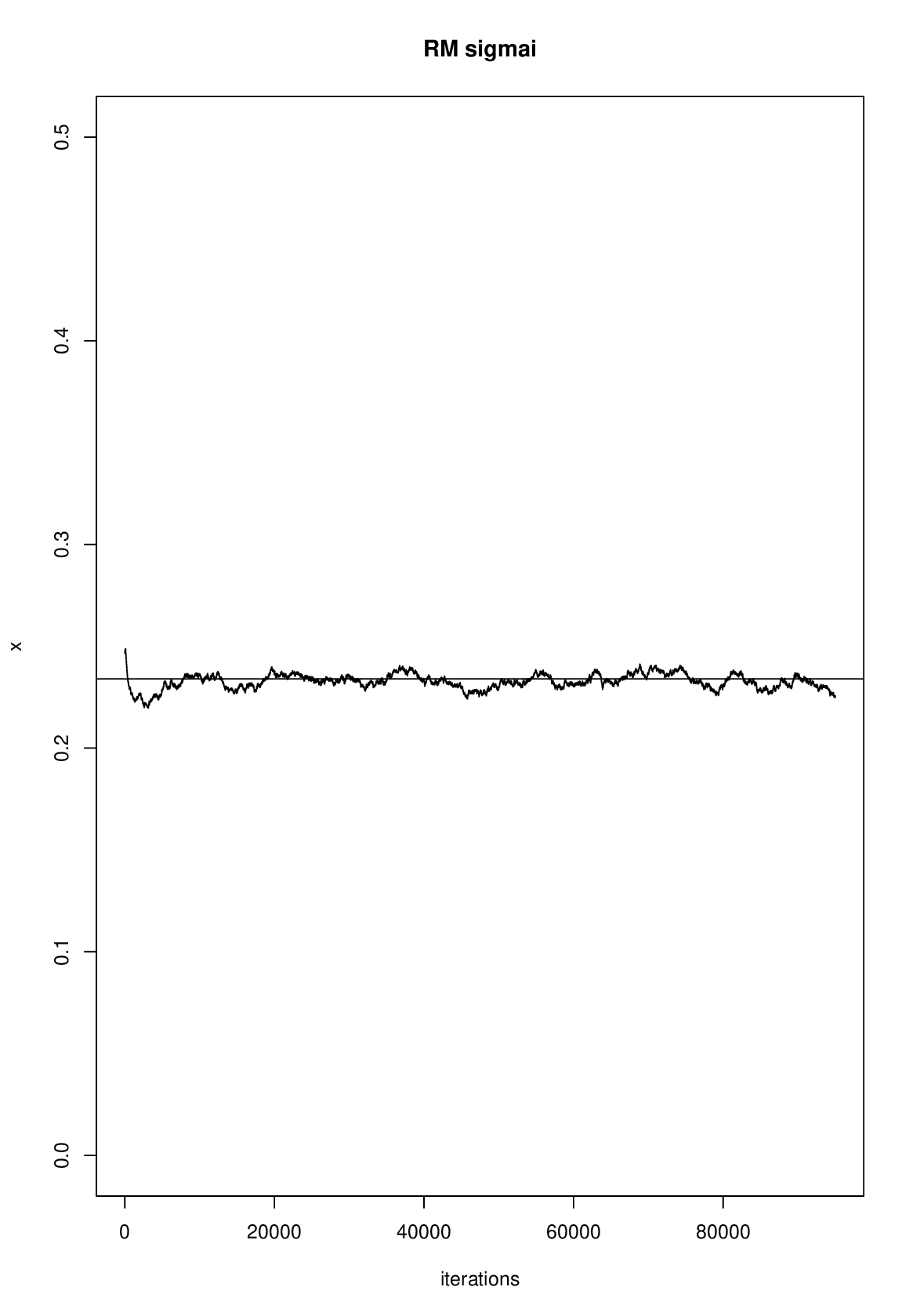} &\psfrag{x}[t][t]{{\tiny OAP}} \includegraphics[width=7cm, height=5cm]{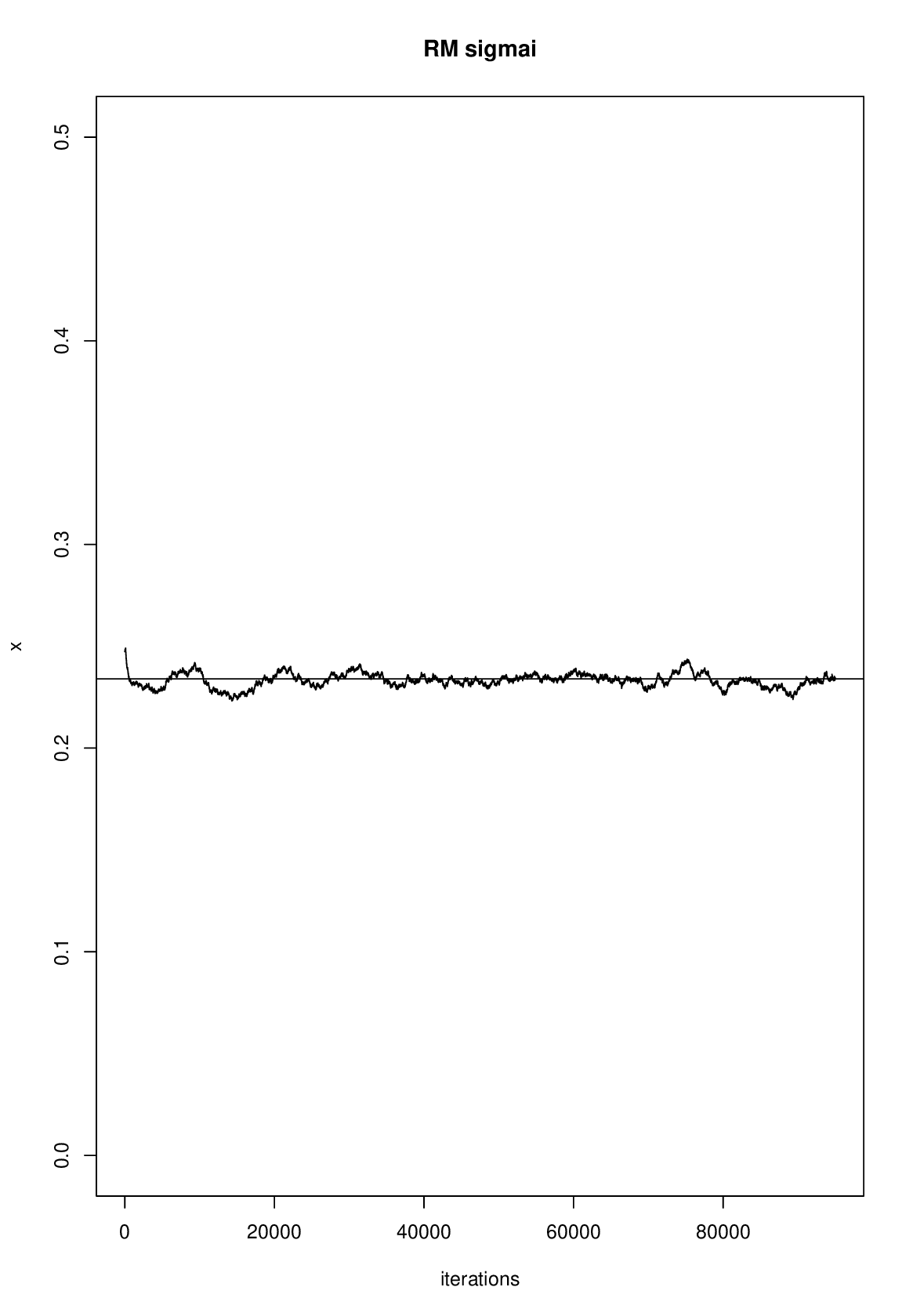}
\end{tabular}
\caption{Trace-plots of $\sigma^2$ (top panels) and the corresponding running acceptance rates (bottom panels) over the previous 500 iterations, using the RM method. Left (Right) panels display results for the better- (ill-) conditioned cases.
}\label{figure5}
\end{center}
\end{figure}

Figure \ref{figure5} shows the ability of the RM method to quickly find the value of $\sigma^2$ that gives the desired OAP for both better- (left panels) and ill-conditioned (right panels) cases. The top panels illustrate the estimates of $(\sigma^*)^2$ for one of the chains from Table \ref{table2};
the horizontal lines correspond to the optimal value, $\sigma =2.38/m^{1/2}$.
The bottom panels display the corresponding running acceptance rates over a window of
500 iterations of the chain. In each chain the estimate of $\sigma^2$ stabilized within about 3000 iterations; thereafter the acceptance rate was close to the target of 0.234. The diagrams illustrate that the RM method performed well with these high-dimensional target distributions, even in the ill-conditioned case.

\subsection{Metropolis-Hastings within Gibbs: Respiratory Infection in Children}
\label{ex:indon}

We apply our adaptive MCMC algorithm to an example involving
respiratory infection in Indonesian children (\shortciteNP{diggle+lz95}; \citeNP{lin+c01}). The data contains longitudinal
measurements on 275 Indonesian children, where the indicator for respiratory
infection is the binary response. The covariates include age, height,
indicators for vitamin A deficiency, gender, stunting and visit numbers
(one to six).
Previous analyses have shown the effect of age of the child to be non-linear,
and so we use a Bayesian logistic additive mixed model of the form
\begin{equation}
\label{eqn:resp}
\mbox{logit}\{P(\mbox{respiratory infection}_{ij}=1)\}=\beta_0+U_i+\bbeta^T\bX_{ij}+f(\mbox{age}_{ij})
\end{equation}
for $1\leq i \leq 275$ children and $1\leq j \leq n_i$ repeated measures within a child.
The random child effect is $U_i \overset{\mbox{\tiny ind.}}{\sim} N(0,\sigma^2_U)$,
$\bX_{ij}$ is the measurment
vector of the remaining 11 covariates, and $f$
is modelled using penalized splines with spline basis coefficients
$u_k \overset{\mbox{\tiny ind.}}{\sim} N(0,\sigma^2_u)$.

We follow \shortciteN{zhao+scw06} and \shortciteN{fan+lw08} and apply hierarchical centering
to the random effects. All continuous covariates are standardised
so that the choice of hyperparameters
can be scale independent. Radial cubic basis functions are used to fit
the covariate age
\[f(\mbox{age})=\beta_0+\beta_1\mbox{age}+\bZ_{\mbox{\tiny age}}\bu\]
where
\[
\bZ_{\mbox{\tiny age}}=[|\underset{1\leq \kappa \leq K}{\mbox{age}-\kappa_k}|^3]
[|\underset{1\leq k, k'\leq K}{\kappa_k-\kappa_{k'}}|^3]^{-1/2} \quad
\mbox{and} \quad \bu \sim N(0,\sigma^2_u\bI)
\]
with $\kappa_k=(\frac{k+1}{K+2})$th quantile of the unique predictor values,
where $K$ is chosen to be $20$.
We use a vague prior for the fixed effects. For both variance components, we use an
inverse gamma prior with equal scale and shape parameters. Previous Bayesian
analyses (e.g., \shortciteNP{zhao+scw06}) showed this to be a robust choice for scale (and shape) parameter values of
0.01 and larger.

We consider two variants of Metropolis-Hastings within Gibbs samplers.
The first (a full-conditional approach) systematically cycles through all 306 parameters, using a RWMH algorithm with proposal distribution $N(x^{t}_{i-1}, \sigma_t^2)$, for each $t=1,\ldots 306$, and
implementing separate Robbins-Monro searches with the aim of finding values of $\sigma _t ^2 $ that give acceptance probabilities of 0.44.
The second scheme (a block-conditional approach), block-updates the eleven $\bbeta$ parameters
and the twenty knot components $\{\kappa_k\}$  respectively via the multivariate Gaussian proposal distributions $N(\bx^{(c)}_{i-1}, \sigma_{(c)}^2\bA_c)$ and $N(\bx^{(k)}_{i-1}, \sigma_{(k)}^2\bA_k)$,
where the superscripts $c$ and $k$ refer to the coefficient component and knot component of $\bx$.
The desired acceptance rate of the associated Robbins Monro searches is 0.234 in each case. The remaining 275 parameters are updated using the full-conditional approach as before.
In both schemes the two variance components are updated using the Gibbs sampler, and all algorithmic conditions follow those
outlined in  Sections \ref{sec:searchuni} and \ref{sec:searchfull}.

The sequential Monte Carlo sampler of \shortciteN{fan+lw08} fits the model (\ref{eqn:resp}) using penalized quasi-likelhood (PQL) \cite{breslow+x95} to obtain an approximate maximum likelihood estimate of the covariance matrix of the 306 parameters.
To provide comparison with the Robbins-Monro method, we implement the full conditional approach but using the diagonal elements of the PQL matrix as (fixed) $\sigma^2$ values. For the block-conditional approach, the appropriate blocks of the PQL matrix were also used instead of $\bA_c$ and $ \bA_k$, where the values of $\sigma_{(c)}^2$ and  $\sigma_{(k)}^2$ are set to the optimal values of 0.51 and 0.28 respectively.  For all samplers, chains of length 10,000 and 50,000 were used for the full- and block-conditional approaches, respectively.

\begin{figure}
\begin{center}
\psfrag{Full conditional}[b][b]{{\tiny Full conditional }}
\psfrag{Block conditional}[b][b]{{\tiny Block conditional }}
\psfrag{OAP}[r][r]{{\tiny OAP }}
\psfrag{sigma}[r][r]{{\tiny $\sigma^2$ }}
\psfrag{RM}[t][t]{{\tiny RM }}
\psfrag{Opt-PQL}[t][t]{{\tiny PQL}}
\psfrag{PQL}[t][t]{{\tiny PQL}}
\begin{tabular}{ll}
\includegraphics[width=7cm, height=3cm]{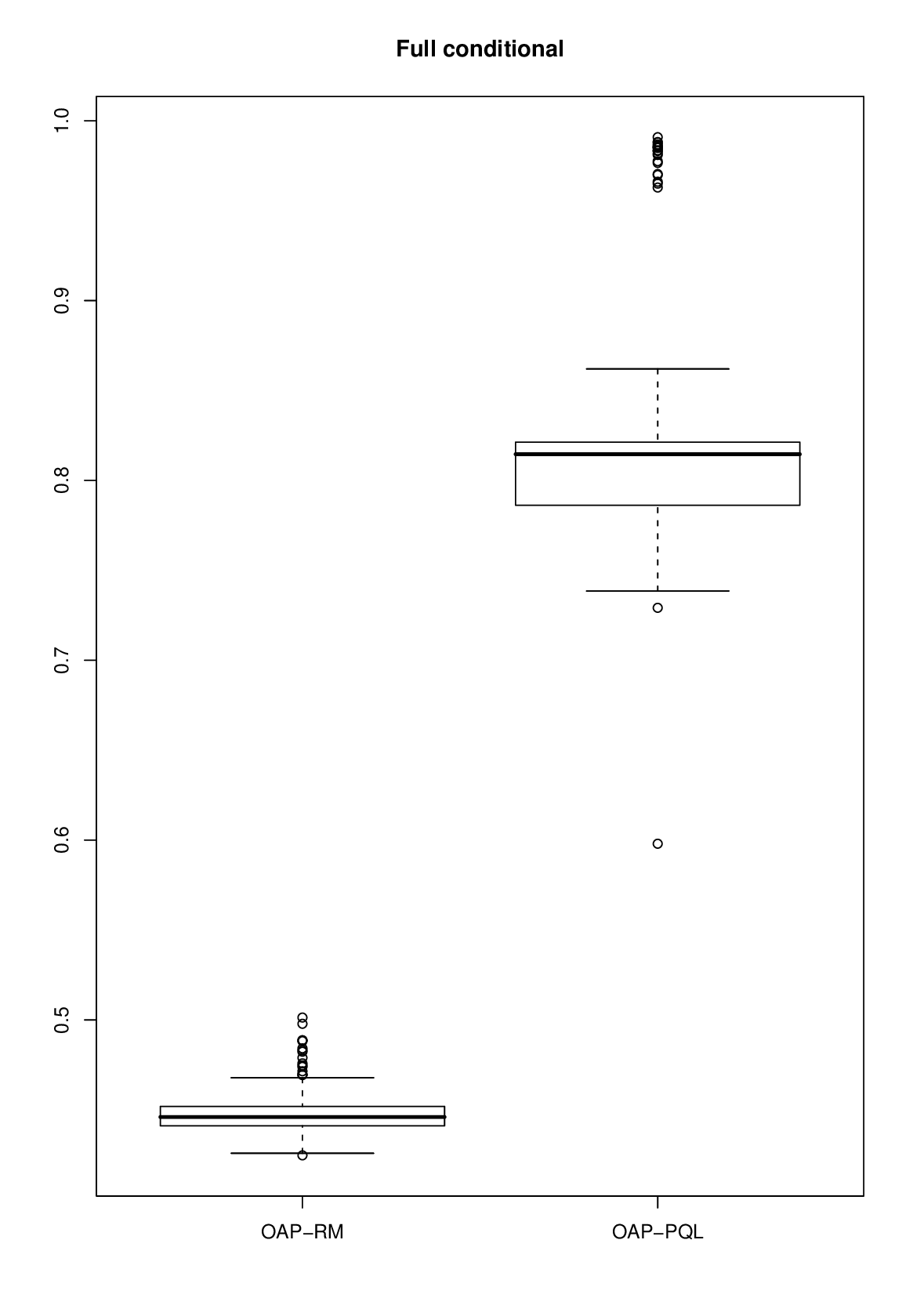} & \includegraphics[width=7cm, height=3cm]{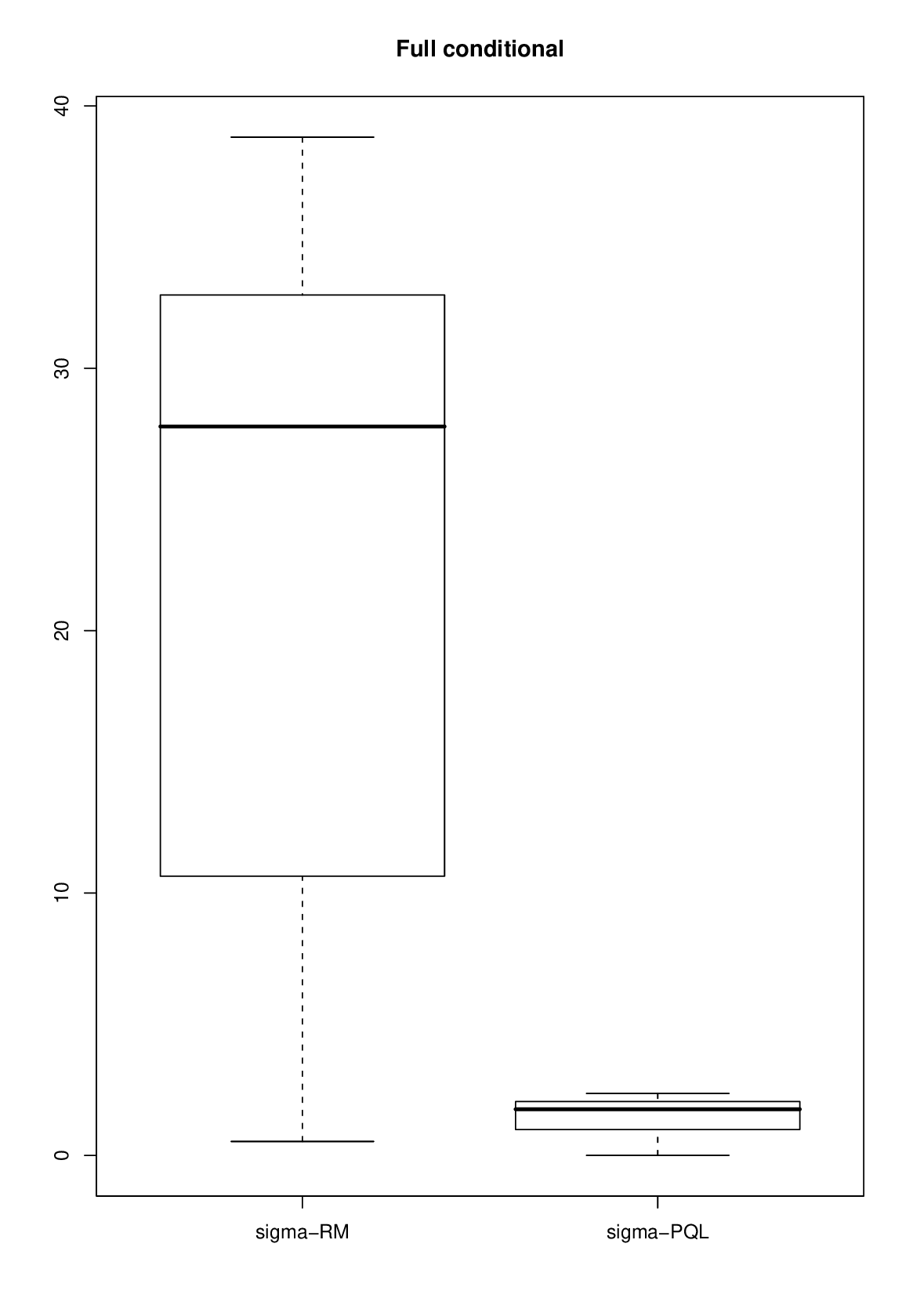} \\
\includegraphics[width=7cm, height=3cm]{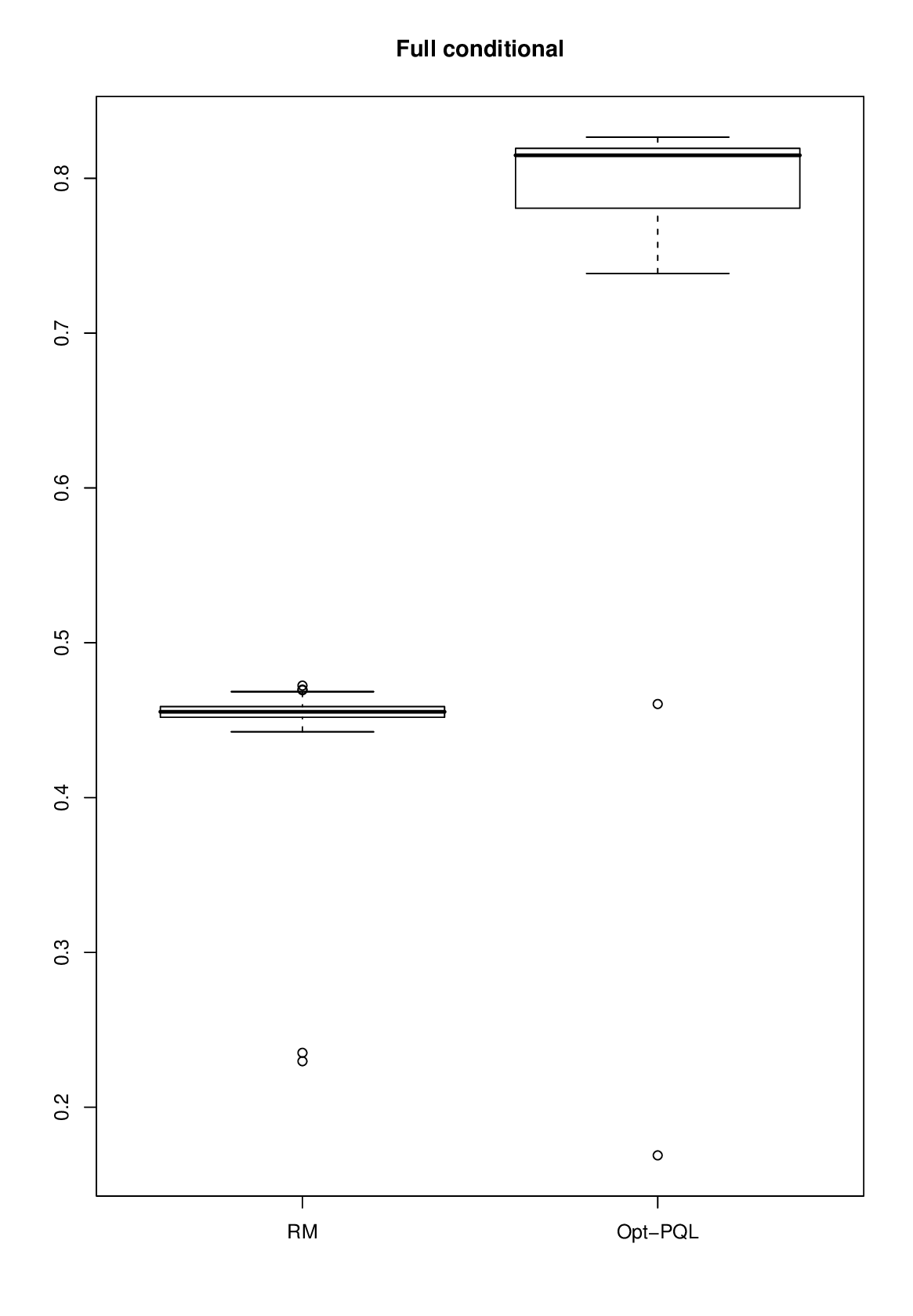} & \includegraphics[width=7cm, height=3cm]{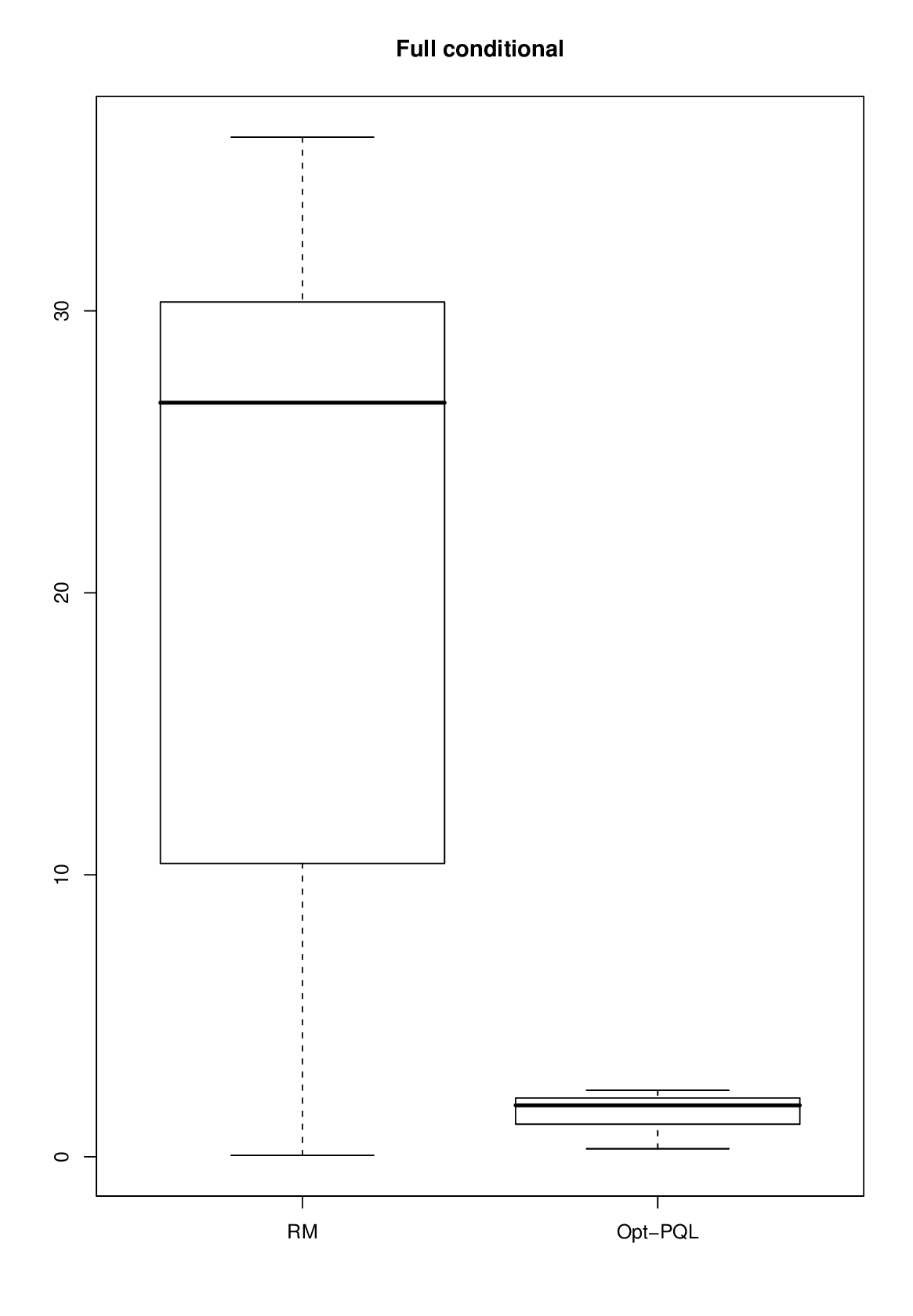} \\
\end{tabular}
\caption{Boxplots of overall acceptance probabilities (left panels) and $\sigma^2$  values (right panels) for all parameters and blocks, based on the
second half of the MCMC sampler output. Top and bottom panels indicate full-conditional and block-conditional updates respectively.}
\label{figure-indon}
\end{center}
\end{figure}

Results for all univariate proposal distributions are summarized in Figure \ref{figure-indon}.
Overall acceptance probabilities (left panels) and the final estimates of $(\sigma^*)^2$ (right panels) for each of the (306 or 275) parameters are illustrated for both the full-conditional (top panels) and block-conditional approach (bottom panels). Within each panel, the left boxplot displays results for the Robbins-Monro method, and the right boxplot the PQL matrix approach.

Both Robbins-Monro samplers led to acceptance probabilities that were
very close to the target of 0.44; the OAP ranged from 0.425 to 0.501 for the 306 proposals run for the full-conditional method and from 0.442 to 0.472 for the 275 univariate proposals of the block-conditional method. To achieve these 
rates the RM search varied $\sigma _t ^2 $ substantially; the highest final mean value (based on the second half of the chain) of $\sigma _t ^2 $ was 38.806 and the lowest final mean value was 0.529. In marked contrast, the method based on PQL
gave OAP values which varied far more than with the RM method. For the full-conditional method, the OAP given by PQL varied from 0.598 to 0.991 -- frequent and substantially different deviations from the ideal target.

The Robbins-Monro searches were also effective with the multivariate proposals. The target OAP was 0.234 and it gave an OAP of 0.235 for the block of eleven $\bbeta$ coefficients and 0.230 for the block of 20 knots $\{\kappa_k\}$.

\section{Discussion}
\label{sec:disc}

This paper exploits the Robbins-Monro process to give a method for automatically tuning the scaling factor of the
Random-Walk Metropolis-Hastings MCMC sampler. The method was implemented in a search algorithm and its effectiveness illustrated. When the random walk covariance matrix was estimated jointly with the scaling factor, our search algorithm for the scaling factor converged so quickly that we had to deliberately slow it down to give time for the covariance matrix to be estimated. This speed of convergence suggests that the algorithm may make a useful component of more complex MCMC methods. For instance, recently proposed MCMC algorithms (\shortciteNP{craiu+ry09}; \citeNP{roberts+r09}) offer regional adaptation in order to improve chain mixing when a global scale parameter is inadequate. These methods require a different proposal distribution to be estimated depending on which region of the parameter space the chain is in currently. Hence the speed with which the proposal distribution is estimated in each region is likely to be an issue, making use of Robbins-Monro searches an attractive approach.

Our search algorithm satisfies the {\it diminishing adaptation criterion} \cite{roberts+r09}, in that changes to the scale parameter vanish as the length of the MCMC chain goes to infinity. An interesting possibility would be to relax this requirement by fixing a (small) minimum value for the size of scale parameter changes. This could allow continuous adaptation of the proposal distribution as the Markov chain moves around the parameter space. Existing theory would need to be extended to determine when the chain converges to its target distribution, but theoretical results might be obtainable as the Robbins-Monro process is itself a Markov chain, so that the Robbins-Monro process and an MCMC chain together still form an MCMC chain.

\subsection*{Acknowledgements}

YF and SAS are supported by the Australian Research Council under the Discovery Project scheme (DP0877432).

\subsection*{Appendix: Proof of Propositions \ref{PropP0} and \ref{PropP1}}

\setcounter{equation}{0}
\renewcommand{\theequation}{A.\arabic{equation}}
\noindent{\em \bf Proof of Proposition \ref{PropP0}}. $\int \mbox{min}( f^{\#}(\by)/f^{\#}(\bx) ,\,1 ) \,  g(\by \, |\, \bx,\sigma) \, d\by  \leq 1$, so
\begin{equation}
\int \int \mbox{min}\left( \frac{f^{\#}(\by)}{f^{\#}(\bx) },\,1 \right) \,  \bx ' \bx \, g(\by \, |\, \bx,\sigma) \, f(\bx) \, d\by \, d\bx \leq \int \bx ' \bx \, f(\bx) \,  d\bx .
\end{equation}
As $f(.)$ has finite variance, both sides of (A.1) are finite. Similarly, as $g(\by \, |\, \bx,\sigma) = g(\bx \, |\, \by,\sigma)$, we have that $\int \int \mbox{min}( f^{\#}(\by)/f^{\#}(\bx) ,\,1 ) \,  \by ' \by \, g(\by \, |\, \bx,\sigma) \, f(\bx) \, d\by \, d\bx $ is also finite. It follows that $\phi$ in equation (4) is $O(\sigma ^{-3})$.

Let $S(r)$ be an $m$-dimensional sphere of radius $r$, centred at the mean of $f(\, . \,)$. Let $S^c(r) = \Omega - S(r)$ denote its complement. Given any $\epsilon$, choose $r$ such that $\int_{S^c(r)} f(\by) \, d\by < \epsilon$. Then $mp(\sigma) \approx \int_{\Omega} \int_{S(r)} \mbox{min}( f^{\#}(\by)/f^{\#}(\bx) ,\,1 ) \, m \, g(\by \, |\, \bx,\sigma) \, f(\bx) \, d\by \, d\bx $ and
\begin{equation}
\lim_{\sigma \rightarrow \infty} mp(\sigma) \approx \int_{\Omega} \int_{S(r)} \mbox{min}\left( \frac{f^{\#}(\by)}{f^{\#}(\bx) },\,1 \right) \, m \, (2\pi )^{-m/2} \sigma ^{-m} \, |\bA | ^{-1/2} \, f(\bx) \, d\by \, d\bx ,
\end{equation}
since $\lim_{\sigma \rightarrow \infty} g(\by \, |\, \bx,\sigma) \rightarrow (2\pi )^{-m/2} \sigma ^{-m} \, |\bA | ^{-1/2}$ for $\by \in S(r)$. As $mp(\sigma)$ is non-zero for finite $\sigma$, $mp(\sigma)$ is $O(\sigma^{-m})$. Hence, $\lim_{\sigma \rightarrow \infty} \{ \phi - mp(\sigma) /\sigma \} = \lim_{\sigma \rightarrow \infty} \{  - mp(\sigma) /\sigma \} $ if $m=1$, since $\phi$ is $O(\sigma ^{-3})$. Then, from equation (3), $\lim_{\sigma \rightarrow \infty} dp(\sigma)/d\sigma =  - mp(\sigma) /\sigma $ and the proposition follows from $c^*/\sigma^*  = -1/ [\sigma \, dp(\sigma)/d\sigma ] _{\sigma =\sigma^*}$. \hfill $\Box $\\

\noindent{\em \bf Preliminary lemma.} Let $g(. | \, \bx, \, \sigma)$ be a multivariate normal distribution and let $\bx$ and $\blam$ be fixed $m \times 1$ vectors. Define the region $R$ by $R = \{ \by: \, \blam ' (\by -\bx) \leq 0 \} $. Then
\begin{equation}
\int _R \blam ' (\by -\bx) \{ dg(\by | \, \bx, \, \sigma)/d\sigma \} d\by \, = \,
\sigma ^{-1} \int _R \blam ' (\by -\bx) g(\by | \, \bx, \, \sigma) d\by.
\end{equation}
\\

\noindent{\em \bf Proof of lemma.} For $i=1,\ldots , m$ and $j=1,\ldots , m$, let $b_{ij}$ be the $(i, \, j)$ element of $\bA ^{-1}$ and let $x_i$, $y_i$ and $\lambda _i$ denote the $i$th elements of $\bx$, $\by$ and $\blam $. Suppose $\lambda _i >0$ and let $h(i) = x_i - \sum _{j: j \neq i } \lambda _j (y_j -x_j )/\lambda _i $. Integrating by parts, $\int_{-\infty}^{h(i)} \{ \blam ' (\by - \bx) \} \{ \sigma^{-3} \sum _j b_{ij} (y_i -x_i ) (y_j -x_j ) \} .g(\by | \, \bx, \sigma ) \, dy_i = [ \{ \blam ' (\by - \bx) \} \{ -\sigma^{-1} (y_i -x_i ) \} .g(\by | \, \bx, \sigma ) ] _{y_i = -\infty} ^{h(i)} \, + \int_{-\infty}^{h(i)} \sigma ^{-1} \{ \blam ' (\by - \bx) +\lambda_i (y_i - x_i ) \} .g(\by | \, \bx, \sigma ) \, dy_i = 0\,+\, \int_{-\infty}^{h(i)} \sigma ^{-1} \{ \blam ' (\by - \bx) +\lambda_i (y_i - x_i ) \} .g(\by | \, \bx, \sigma ) \, dy_i .$ Thus
\[
\int_{R} \{ \blam ' (\by - \bx) \} \{ \sigma^{-3} \sum _j b_{ij} (y_i -x_i ) (y_j -x_j ) \} .g(\by | \, \bx, \sigma ) \, d\by \vspace{-.2in}
\]
\begin{equation}
=  \int_{R} \sigma ^{-1} \{ \blam ' (\by - \bx) +\lambda_i (y_i - x_i ) \} .g(\by | \, \bx, \sigma ) \, d\by .
\end{equation}
Equation (A.4) also holds for $\lambda _i \leq 0$, so $ \int_{R} \{ \blam ' (\by - \bx) \} \{ \sigma^{-3} \sum _i \sum _j b_{ij} (y_i -x_i ) (y_j -x_j ) \}$ $ .g(\by | \, \bx, \sigma ) \, d\by = \int_{R} (m+1) \sigma ^{-1} \{ \blam ' (\by - \bx) \} .g(\by | \, \bx, \sigma ) \, d\by .$ Since $ dg(\by | \, \bx, \sigma ) / d\sigma =$ $ \{ \sigma^{-3} \sum _i \sum _j $ $b_{ij} (y_i -x_i ) (y_j -x_j ) - m\sigma^{-1} \} .g(\by | \, \bx, \sigma )$, the lemma follows.$\, \, \,\hfill \Box$\vspace{.1in}\\

\noindent{\em \bf Proof of Proposition \ref{PropP1}.} As $p(\sigma) \rightarrow 1$, $\sigma \rightarrow 0$. Then $g(\by|\, \bx, \sigma )\rightarrow 0$ except when $|| \, \by-\bx|| \rightarrow 0$. Let $\bx $ be fixed. For small $|| \, \by-\bx||$, we may put $f(\by) \approx f(\bx) +\blam ' (\by-\bx) f(\bx)$, where $\blam$ does not depend on $\by$. Then $f^{\#}(\by)/f^{\#}(\bx) \leq 1$ if and only if $\blam ' (\by -\bx) \leq 0$. Let $R^c = \{ \by: \, \blam ' (\by -\bx) > 0 \} $ so that $R^c$ is the complement of $R$. As $\sigma \rightarrow 0$,
\[
\int \mbox{min} \left(\frac{f^{\#}(\by)}{f^{\#}(\bx)} , \, 1 \right) .\, g(\by \, |\, \bx,\sigma)\,  d\by
 \approx \int_{R^c} g(\by|\, \bx,\sigma) \, d\by
\, + \, \int_R (1+\blam ' (\by -\bx) )g(\by|\, \bx,\sigma) \, d\by
\]
\begin{equation}
=  1 \, + \, \int_R \blam ' (\by -\bx) g(\by|\, \bx,\sigma) \, d\by.
\end{equation}
Differentiating equation (A.5), as $\sigma \rightarrow 0$,
\[
\int \mbox{min} \left(\frac{f^{\#}(\by)}{f^{\#}(\bx)} , \, 1 \right) .\, \frac{dg(\by \, |\, \bx,\sigma)}{d\sigma} \,  d\by
 \approx \int_R \blam ' (\by -\bx) )\, \frac{dg(\by \, |\, \bx,\sigma)}{d\sigma} \,  d\by
\]
\begin{equation}
= \sigma ^{-1} \int _R \blam ' (\by -\bx) g(\by | \, \bx, \, \sigma) d\by ,
\end{equation}
from the lemma. From equations (A.5) and (A.6),
\[
\int_{R} \mbox{min} \left( \frac{f^{\#}(\by)}{f^{\#}(\bx)}, \, 1  \right)   .\, g(\by \, |\, \bx,\sigma)\,  d\by \, \approx 1 \,+ \,  \sigma \int_R \mbox{min} \left( \frac{f^{\#}(\by)}{f^{\#}(\bx)}, \, 1  \right)
 .\, \frac{dg(\by \, |\, \bx,\sigma )}{d\sigma  } \,  d\by .
\]
Now, from Equation \ref{eqn:p},
\[
\int \int \mbox{min} \left( \frac{f^{\#}(\by)}{f^{\#}(\bx)}, \, 1  \right)   .\, g(\by \, |\, \bx,\sigma)\, f(\bx) \, d\by \, d\bx \, = \, p(\sigma)
\]
and, from Equation \ref{dpds},
\[
\int \int   \mbox{min} \left( \frac{f^{\#}(\by)}{f^{\#}(\bx)}, \, 1  \right)
 .\, \frac{dg(\by \, |\, \bx,\sigma )}{d\sigma  }  \, f(\bx) \, d\by \, d\bx \,
= \, \frac{dp(\sigma)}{d\sigma}.
\]
Hence, as $p(\sigma) \rightarrow 1$, $p(\sigma ) \approx 1 +\sigma \{ dp(\sigma)/d\sigma \} $, so $c^* = -1 / [ dp(\sigma) / d\sigma ] _{\sigma =\sigma^*} \approx \sigma ^* /(1-p^* ). \, \, \, \Box$

\bibliographystyle{RM}
\bibliography{RM}

\end{document}